\documentclass[aps,pra,twocolumn,superscriptaddress,amsmath,amssymb,showpacs,notitlepage]{revtex4-1}
\pdfoutput=1
\usepackage[caption=false]{subfig}
\usepackage{algorithm}
\usepackage{color}
\usepackage{algorithmic}
\usepackage{amsmath}
\usepackage{amssymb}
\usepackage{amsthm}
\usepackage{bm}
\usepackage{thmtools}

\usepackage{array}
\usepackage{url}
\usepackage{hyperref}
\usepackage{graphicx}
\usepackage{bibentry}
\usepackage{microtype}

\providecommand{\U}[1]{\protect\rule{.1in}{.1in}}
\newtheorem*{theorem*}{Theorem}

\newcommand{\ket}[1]{\left| #1 \right>} 

\begin{document}
\title{Non-producibility of arbitrary non-Gaussian states using zero-mean Gaussian states and partial photon number resolving detection}
\author{Christos N. Gagatsos}
\affiliation{James C. Wyant College of Optical Sciences, University of Arizona, 1630 E. University Blvd., Tucson, AZ 85721}
\author{Saikat Guha}
\affiliation{James C. Wyant College of Optical Sciences, University of Arizona, 1630 E. University Blvd., Tucson, AZ 85721}
\affiliation{Department of Electrical and Computer Engineering, University of Arizona, 1230 E Speedway Blvd., Tucson, AZ 85721}

\begin{abstract}
Gaussian states and measurements collectively are not powerful-enough resources for quantum computing, as any Gaussian dynamics can be simulated efficiently, classically. However, it is known that any one non-Gaussian resource---either a state, a unitary operation, or a measurement---together with Gaussian unitaries, makes for universal quantum resources. Photon number resolving (PNR) detection, a readily-realizable non-Gaussian measurement, has been a popular tool to try and engineer non-Gaussian states for universal quantum processing. In this paper, we consider PNR detection of a subset of the modes of a zero-mean pure multi-mode Gaussian state as a means to herald a target non-Gaussian state on the undetected modes. This is motivated from the ease of scalable preparation of Gaussian states that have zero mean, using squeezed vacuum and passive linear optics. We calculate upper bounds on the fidelity between the actual heralded state and the target state. We find that this fidelity upper bound is $1/2$ when the target state is a multi-mode coherent cat-basis cluster state, a resource sufficient for universal quantum computing. This proves that there exist non-Gaussian states that are not producible by this method. Our fidelity upper bound is a simple expression that depends only on the target state represented in the photon-number basis, which could be applied to other non-Gaussian states of interest.
\end{abstract}
\maketitle

\section{Introduction}
Production of non-Gaussian quantum states of light, and all-optical realization of non-Gaussian quantum unitary operations, are critical for most applications of photonic quantum information processing, e.g., universal photonic quantum computation~\cite{Pant2019}, quantum-enhanced receivers for optical communications~\cite{Guha2011,Sabuncu2010}, all-optical quantum repeaters for long-distance entanglement distribution~\cite{Lo2015, Pant2015,He2020}, and quantum-enhanced optical sensing~\cite{Tan2008, Guha2009, Zhuang2017, Dowling2008, Humphreys2013,Pirandola2018}. 

Gaussian states and Gaussian unitaries, produced by the action of linear and quadratic Hamiltonians on the vacuum state, have efficient and complete mathematical representations~\cite{Reck1994, Clements2016, Braun2005}. Non-Gaussian states is a vast set---it consists of states generated via the action, on the multi-mode vacuum state, of a unitary with Hamiltonian that is a third or higher-order polynomial in the field operators. Therefore, non-Gaussian states are inherently under-explored and their general representations less-understood. 

Deterministic realization of non-Gaussian unitary operations, such as the self-Kerr gate~\cite{KLM2001, Kok2007} and the cubic-phase gate~\cite{GKP2001} is near impossible at optical frequencies~\cite{Shapiro2006}. The extreme resource inefficiency resulting from this deficiency, combined with the fact that Gaussian states and Gaussian unitaries are a classically-simulable resource~\cite{Niset2009}, have kept all-photonic quantum computing from being pursued as one of the top contenders for quantum computing, for decades since their invention, despite their obvious importance in optical communications and sensing applications, and it not requiring quantum transduction for networking far-flung quantum processors---a major benefit unique to photonic quantum encodings.

Recent advances in discrete variable (DV), i.e., single-photon-qubit based, quantum computing~\cite{KLM2001} have revealed that deterministic production of even small non-Gaussian resource states (e.g., a $3$-photon-entangled GHZ state) can enable resource-efficient universal photonic quantum computing, despite two-qubit gates being inherently probabilistic~\cite{Mercedes2015, Pant2019}. However, a systematic understanding of efficient production of even such simple non-Gaussian states as GHZ states and realization of simple two-qubit non-Gaussian measurements (e.g., Bell-state measurements) required for DV quantum computing has proven extremely difficult~\cite{Frederic2018}.

A major attraction of continuous variable (CV) quantum computing~\cite{GKP2001} is that large Gaussian entangled (cluster) states \cite{Pfister2021} can be produced experimentally in a one-shot deterministic fashion~\cite{Pfister2011, Furusawa2013}. Further, CV qubit states, such as the GKP qubit is known to be the most loss-resilient encoding of the qubit in a bosonic mode~\cite{Noh2019-ak} that admit deterministic Clifford gates using Gaussian unitaries, and there are native CV quantum codes to correct for loss errors~\cite{Lassen2010}. Since Gaussian states are not universal~\cite{Niset2009}, one needs a non-Gaussian operation to enable universal quantum computing~\cite{Braun1999}. Experimentally, the most readily-available non-Gaussian resource is photon number resolving (PNR) detection~\cite{NISTTES}. One common modality in which a PNR detector can be used to probabilistically engineer non-Gaussian states is {\em photon subtraction}~\cite{Jaromir2005,Grangier2009}, which also is known to increase entanglement~\cite{Grangier2007}. Photon subtraction from multi-mode Gaussian states has been achieved experimentally~\cite{Ra2017, Ra2019, Takahashi2008} and several theoretical aspects of photon subtraction have been studied~\cite{Treps2016, Brouri2009, Marek2008, Barnett2018, Walschaers2018,Arzani2019,Mattia2017,Ferrini2017}. In this paper, we consider a more general and simpler to describe set-up, namely \emph{partial PNR} detection~\cite{Su2019,Gagatsos2019}, i.e., employing PNR detection on a subset of the modes of a multi-mode Gaussian state, to herald the undetected modes in a desired state, conditioned on the PNR detectors' click pattern on the detected modes. The heralded state is non-Gaussian unless all PNR detectors register zero photon clicks. This is because the projection on vacuum is a Gaussian operation and therefore will not impart any non-Gaussianity to the heralded state). We point out that partial measurements has been used before in different contexts, e.g., to realize minimal-disturbance measurements experimentally~\cite{sabuncu2007}. 

The most general multimode Gaussian state is described by a covariance matrix and a coherent displacement vector. Zero-mean Gaussian states are those whose displacement vector, or the mean field amplitude, is zero. A general $K$-mode Gaussian state can be produced by passing $K$ displaced-squeezed states through a linear-optical unitary transformation, which in turn admits a systematic design in terms of $K(K-1)$ 50-50 beamsplitters and an equal number of unspecified phase elements~\cite{Reck1994,Clements2016}. Experimentally, the most challenging part in the above is preparing a {\em displaced} squeezed state. Preparation of squeezed vacuum state, or two-mode squeezed vacuum state of light---both of which are zero-mean Gaussian states---on the other hand, is routinely performed using spontaneous parametric downconversion (SPDC), e.g., using a $\chi^{(2)}$ non-linear medium. Recent experiments have demonstrated on-chip squeezed-vacuum generation~\cite{Zhang2021-le,Yang2021-ip}. Further, since fully-programmable linear optical circuits have also been realized on-chip~\cite{Harris2018-vl}, scalable generation of arbitrary multi-mode {\em zero-mean} Gaussian states is well-within the reach of modern technology. This is the reason why we focus in this paper, on evaluating whether arbitrary non-Gaussian states can be prepared just by partial-PNR detection on zero-mean Gaussian states.


This paper is organized as follows: In Sec.~\ref{Sec:PartialPNR} we explain how photon subtraction and addition can be seen as a special case of partial PNR. Said section can be skipped by the experienced reader but it possesses some pedagogical value, and it sets notation. In Sec.~\ref{Sec:Math} we review briefly the mathematical description of partial PNR (see also App.~\ref{App} and~\cite{pizzimenti2021}) and we also provide a simple proof that any pure zero-mean Gaussian state engineered with partial PNR will necessarily give a zero-mean non-Gaussian state. In Sec.~\ref{Sec:Fidelity} we present the fidelity between the heralded state and a given non-Gaussian (target) state. The idea behind the upper bound on said fidelity is based on the Cauchy-Schwartz inequality. In Sec.~\ref{Sec:Product}, we calculate said bound for any product state of single-mode superposition of the binary phase-shift keyed (BPSK) coherent states $|-\gamma\rangle$, $|\gamma\rangle$ and give a few examples. In Sec.~\ref{Sec:Entangled} we calculate fidelity upper bounds for multimode entangled states that are superpositions of multimode coherent states where each mode is from the BPSK constellation. The first special case of an entangled state of this type that we consider is the coherent GHZ state, for which our fidelity upper bound comes out to $1$ (a trivial upper bound). However, for the coherent-cat basis cluster states (CCCS), our fidelity upper bound evaluates to $1/2$, showing such a state cannot be prepared by partial PNR on a zero-mean Gaussian state. Finally, in Sec.~\ref{Sec:Conclusions}, we summarize our results: we discuss how the fidelity upper bound relates to the absence of a non-zero mean (or coherent displacement), we put our findings in context with the literature on non-Gaussian quantum state preparation, present further intuition, and briefly discuss future directions of research.

\section{Partial PNR as a generalization of photon subtraction and addition}\label{Sec:PartialPNR}
Let us consider photon subtraction on the most general single-mode Gaussian pure state: a squeezed-coherent state $|\alpha_1, r_1\rangle$, where $\alpha_1,\ r_1\ \in \mathbb{C}$ are the displacement and squeezing parameters respectively. The state interacts with vacuum on a beamsplitter of transmissivity $\tau \sim 1$. On the reflective (low-transmissivity) output port of the beamsplitter, PNR detection is performed, which registers, say, $n_1$ photons. This heralds the {\em subtraction} of $n_1$ photons from the input state. This conditional photon-subtracted state $|\Phi_{n_1}\rangle$---i.e., the state heralded on the transmitted port of the beamsplitter---is non-Gaussian whenever $n_1\geq 1$. One can write down the probability of detecting $n_1$ photons (and hence producing $|\Phi_{n_1}\rangle$) as a function of $\alpha_1, r_1$ and $\tau$~\cite{Gagatsos2019}. A natural generalization of this setup is to allow for further Gaussian resources by substituting the vacuum state at the other input of the beamsplitter with another single mode Gaussian pure state $|\alpha_2, r_2\rangle$, $\alpha_2,\ r_2\ \in \mathbb{C}$, and then proceed with PNR detection on one output port, and considering the heralded state on the other output port if $n_1$ photons are detected. No matter what figure of merit we might choose on the quality of the heralded state (e.g., {\em fidelity} to some target state), it can only improve, or at worst remain the same as compared to that using photon subtraction. This is because vacuum is a trivial special case of a pure Gaussian state. We term this setup partial PNR detection~\cite{Su2019}: PNR detection on one mode of a two-mode general Gaussian pure state, to seek a desired post-selected non-Gaussian state on the undetected mode.

Next, let us consider photon subtraction on all modes of a $K$-mode pure Gaussian state (which in general is entangled), by coupling the $i$-th mode with vacuum on a beamsplitter of transmissivity $\tau_i,\ i=1,\ldots,K$. If we count all the ancillary vacuum states as input modes, we have a $2K$-mode Gaussian state $K$ modes of which are detected with PNR, resulting in a $K$-mode (generally non-Gaussian) state. An obvious generalization is to consider an $N$-mode Gaussian pure state (where $N$ can be even or odd) and to apply PNR detection on $N-M$ modes, resulting in an $M$-mode state $|\Phi_{n_{M+1},\dots,n_N}\rangle$, conditioned on the PNR pattern $(n_{M+1}\dots,n_N)$.

Let us note that partial PNR detection schemes incorporates multiple photon addition as well. Photon addition is modeled utilizing a beam splitter whose upper input is the state $|\Psi\rangle$ (or a mode of the state) which will undergo photon addition while in the beam splitter's lower input port a Fock state $|n\rangle$ is injected. A PNR detector is applied on the lower output port which if heralds $m \leq n$ photons, then photons have been added to the state $|\Psi\rangle$, resulting to a state $|\Phi\rangle$ which in general will be non-Gaussian. To produce a Fock state $|n\rangle$, i.e., the input to the lower port of the beam splitter, one can consider a two-mode squeezed vacuum state (TMSV) whose one mode is detected using a PNR detector. Then, this Fock state can be used for the photon addition task. Equivalently, one can include several TMSV states  as part of a general Gaussian state and all the PNR detectors to be included in the very last step of the state engineering protocol.

\section{Mathematical description of partial PNR on zero-mean Gaussian states}\label{Sec:Math}
Let us consider a zero-mean $N$-mode Gaussian (in general entangled) pure state $|\Psi\rangle$, prepared by mixing $N$ single-mode squeezed vacuum states of squeezing parameters $r_i$, $1 \le i \le N$, in a linear optical $N$-mode unitary operation $U$. We consider partial PNR detection on $N-M$ of those modes. We proved that (see \cite{Gagatsos2019} and Section II of \cite{pizzimenti2021}, also included as App. \ref{App} of this paper in uniform notation, for the benefit of the reader), conditioned on the PNR detection pattern $(n_{M+1}\dots,n_N)$, the $M$-mode heralded state $|\Phi_{n_{M+1},\dots,n_N}\rangle\equiv |\Phi\rangle$ can be written in the Fock basis as: 
\begin{eqnarray}
\label{eqq:Phi}|\Phi\rangle = \sum_{n_1,\ldots,n_M=0}^{\infty}c_{n_1 \ldots n_M} |n_1 \ldots n_M\rangle,
\end{eqnarray}
where,
\begin{eqnarray}
\label{eqq:FockCoef}c_{n_1 \ldots n_M} = \frac{\mathcal{I}_{n_1 \ldots n_M n_{M+1}\ldots n_N}}{\sqrt{P} \prod\limits_{i=1}^{N} \sqrt{n_i!2^{n_i} \cosh r_i}}.
\end{eqnarray}
The probability of obtaining the PNR pattern $(n_{M+1},\ldots,n_N)$ is given by,
\begin{eqnarray}
	\label{eqq:GBSprob3} P= \sum_{n_1,\ldots,n_M=0}^\infty\frac{\big|\mathcal{I}_{n_1\ldots n_N}\big|^2}{ \prod\limits_{i=1}^{N} n_i!2^{n_i} \cosh r_i},
\end{eqnarray}
and
\begin{eqnarray}
\label{eqq:Hafnian}	\mathcal{I}_{n_1 \ldots n_N} = \left\{
\begin{array}{ll}
0&\sum_{i=1}^N n_i=\textrm{odd},\\
&\\
\textrm{Hf}\left(\sigma\right)&\sum_{i=1}^N n_i= \textrm{even},
\end{array}
\right.
\end{eqnarray}
where the loop-Hafnian in Eq. \eqref{eqq:Hafnian} is evaluated for a matrix $\sigma$, whose elements are given in the Appendix, in Eq. \eqref{eq:FandHinv2}. Per Eq. \eqref{eq:Hinv}, $\sigma$ is directly related to the covariance matrix of the $Q$ function of the state $|\Psi\rangle$ on which the partial PNR is performed. As such, the matrix elements of $\sigma$ are functions of the squeezing parameters $r_i$ and the entries of $U$.

Without loss of generality, we set the modes undergoing PNR to be the `last' $N-M$ modes of $|\Psi\rangle$. Further, we consider $r_i>0$ to be real-valued, or equivalently, all phases are pushed into the passive interferometer $U$ that entangles the squeezed vacuum states to create the resource Gaussian state $|\Psi\rangle$.

For the main result of this paper, we will not need to invoke the explicit dependence of $\sigma$ on $|\Psi\rangle$, through the squeezing parameters $r_i$ and the parameters of the entangling passive linear-optical unitary $U$. The property of importance to us will be the parity of the PNR detector's pattern in Eq. \eqref{eqq:Hafnian}.

Let us now prove that if the $N$-mode Gaussian resource state $|\Psi\rangle$ is zero-mean, the heralded $M$-mode conditional state $|\Phi\rangle$ is also zero mean. In other words, the mean field amplitude of $|\Phi\rangle$ is zero, i.e., $\langle \Phi | \hat{a}_i |\Phi\rangle \equiv \langle \hat{a}_i \rangle_\Phi = 0$ for all $i \in \left\{1,\ldots,M\right\}$. Expressing this condition in the Fock basis, we have:
\begin{eqnarray}
\nonumber \langle \hat{a}_i \rangle_\Phi = \sum_{n_1,\ldots,n_M=0}^{\infty} \sqrt{n_i+1} c_{n_1\ldots n_i \ldots n_M} c_{n_1 \ldots n_i+1 \ldots n_M}^*.
\end{eqnarray}
As per Eqs. \eqref{eqq:FockCoef} and \eqref{eqq:Hafnian}, the coefficients $c_{n_1\ldots n_i \ldots n_M}$ (and hence their complex conjugates) are non-zero only if $n_1+\ldots+n_i+\ldots+n_N=\text{even}$. Therefore, for each non-zero term in the sum above, $n_1+\ldots+n_i+\ldots+n_N=\text{even}$, and hence $n_1+\ldots+n_i+1+\ldots+n_N=\text{odd}$. Hence, $c_{n_1 \ldots n_i+1 \ldots n_M}^*=0$, rendering every term in the sum to be zero. Therefore, $\langle \hat{a}_i \rangle_\Phi=0$.


\section{Fidelity upper bound on the conditional state}\label{Sec:Fidelity}
In Sec. \ref{Sec:Math} we proved that a zero-mean Gaussian pure state under partial PNR will necessarily give a zero-mean conditional state on the unmeasured modes. Therefore, it is natural to anticipate that any non-Gaussian target state with an arbitrary non-zero mean-field would not have a fidelity arbitrarily close to $1$ with a non-Gaussian state engineered using partial PNR on zero-mean Gaussian states. However, the question of what the highest said fidelity can be, remains open. In this section, we provide a general recipe to find an upper bound to the fidelity for any target state. In subsequent sections, we will apply this technique to evaluate our fidelity upper bound on specific non-Gaussian states of interest.


The fidelity $\mathcal{F}=|\langle \Phi_t| \Phi \rangle|^2$ between the conditional state $|\Phi\rangle$ of Eq. \eqref{eqq:Phi} and a non-Gaussian target state $|\Phi_t\rangle=\sum_{n_1,\ldots,n_M=0}^{\infty} d_{n_1\ldots n_M}|n_1\ldots n_M\rangle$ reads,
\begin{eqnarray}
	\label{eqq:Fidelity1} \mathcal{F}= \Big| \sum_{n_1,\ldots,n_M=0}^\infty c_{n_1\ldots n_M }^* d_{n_1\ldots n_M} \Big|^2.
\end{eqnarray}
It is apparent that if we use the Cauchy-Schwartz inequality on Eq. \eqref{eqq:Fidelity1}, we will get $\mathcal{F}\leq 1$. However, we will see that under the constraint (Eqs. \eqref{eqq:FockCoef} and \eqref{eqq:Hafnian}) $n_1+\ldots+n_N=\text{even}$, the Cauchy-Schwartz inequality gives a non-trivial upper bound. Then $n_1+\ldots+n_M=\text{even}$ if the summation of the PNR pattern $(n_{M+1},\ldots,n_N)$ is even, and $n_1+\ldots+n_M=\text{odd}$ if the summation of the PNR pattern $(n_{M+1},\ldots,n_N)$ is odd. Therefore, we rewrite Eq. \eqref{eqq:Fidelity1} as,
\begin{eqnarray}
\label{eqq:Fidelity2}	\mathcal{F} = \left\{
\begin{array}{ll}
\mathcal{F}_{\text{even}}, &\sum_{i=n_{M+1}}^N n_i=\text{even},\\
&\\
\mathcal{F}_{\text{odd}}, &\sum_{i=n_{M+1}}^N n_i= \textrm{odd},
\end{array}
\right.
\end{eqnarray}
where,
\begin{eqnarray}
\label{eqq:FidelityEven}	\mathcal{F}_{\text{even}}  &=& \Bigg| \sum_{\substack{n_1,\ldots,n_M=0  \\ n_1+\ldots+n_M=\text{even}}}^\infty c_{n_1\ldots n_M }^* d_{n_1\ldots n_M} \Bigg|^2,\\
\label{eqq:FidelityOdd}\mathcal{F}_{\text{odd}}&=&\Bigg| \sum_{\substack{n_1,\ldots,n_M=0  \\ n_1+\ldots+n_M=\text{odd}}}^\infty c_{n_1\ldots n_M }^* d_{n_1\ldots n_M} \Bigg|^2.
\end{eqnarray}
Let us consider the case where $n_1+\ldots+n_M=\text{even}$. Then $\mathcal{F}_{\text{odd}}=0$ and  we can use the Cauchy-Schwartz inequality to get,
\begin{eqnarray}
\nonumber	\mathcal{F}_{\text{even}} &\leq& \sum_{\substack{n_1,\ldots,n_M=0  \\ n_1+\ldots+n_M=\text{even}}}^\infty |c_{n_1\ldots n_M }|^2 \\
\nonumber \times &&\sum_{\substack{n_1,\ldots,n_M=0  \\ n_1+\ldots+n_M=\text{even}}}^\infty |d_{n_1\ldots n_M }|^2.
\end{eqnarray}  
 Finally, exploiting the fact that the state $|\Phi\rangle$ has non-zero coefficients under the constraint $n_1+\ldots+n_M=\text{even}$, we write,
 \begin{eqnarray}
\sum_{\substack{n_1,\ldots,n_M=0  \\ n_1+\ldots+n_M=\text{even}}}^\infty |c_{n_1\ldots n_M }|^2=1,
 \end{eqnarray}
and we get,
\begin{eqnarray}
\label{eqq:ueven}\mathcal{F}_{\text{even}}\leq \sum_{\substack{n_1,\ldots,n_M=0  \\ n_1+\ldots+n_M=\text{even}}}^\infty |d_{n_1\ldots n_M }|^2=u_{\text{even}}.
\end{eqnarray}
Similarly, for the complementary case where $n_1+\ldots+n_M=\text{odd}$, we have that $\mathcal{F}_\text{even}=0$ and
\begin{eqnarray}
\label{eqq:uodd}	\mathcal{F}_{\text{odd}}\leq \sum_{\substack{n_1,\ldots,n_M=0  \\ n_1+\ldots+n_M=\text{odd}}}^\infty |d_{n_1\ldots n_M }|^2=u_{\text{odd}}.
\end{eqnarray} 
Four observations are necessary here. First, we observe that $0\leq u_{\text{even}} \leq 1$ and $0\leq u_{\text{odd}} \leq 1$ and both bounds depend only on the target state, therefore they are easy to compute.
Second, the non-Gaussian target state is normalized, therefore,
\begin{eqnarray}
\label{eqq:1minusuEven}	u_{\text{odd}}=1-u_{\text{even}}.
\end{eqnarray}
It is possible that for the desired non-Gaussian target state, $u_{\text{even}}$ and $u_{\text{odd}}$ to be unequal. In that case, we will use as upper bound the larger among the two, and herald on the PNR pattern whose parity corresponds to that of the higher upper bound. Third, let us note that we view fidelity as necessary criterion for successful non-Gaussian state engineering. For example, a coherent cat state $N_0^{-1}(|\gamma\rangle+|-\gamma\rangle)$ (where $|\pm \gamma\rangle$ is a coherent state and $N_0$ is normalization) can have high fidelity with vacuum for small, albeit non-zero, $\gamma$ amplitude. However, vacuum and small coherent cat states are inherently different. On the other hand, if one derives a low enough upper bound for the fidelity, then the impossibility of producing the state under consideration is certain. Last, we observe that assuming a zero-mean Gaussian resource state, resulted to imposing a specific parity on the PNR pattern. The question now is how this parity constraint impacts the state engineering performance.

\section{Fidelity upper bounds for coherent cat product states}\label{Sec:Product}
Consider a single mode state $|c\rangle$, which is a superposition of two coherent states $|\pm \gamma\rangle$,
\begin{eqnarray}
\label{eqq:statec}	|c\rangle = b_1 |\gamma\rangle+b_2 |-\gamma\rangle,
\end{eqnarray}
where $b_1,\ b_2\ \in \mathbb{C}$ satisfy,
\begin{eqnarray}
\label{eqq:normc}	(b_1 b_2^*+b_1^* b_2) e^{-2 |\gamma|^2}=1-|b_1|^2-|b_2|^2,
\end{eqnarray}
so that $\langle c|c\rangle = 1$. Let us calculate the upper bounds of Eqs. \eqref{eqq:ueven} and \eqref{eqq:uodd} for the product state $|c\rangle^{\otimes M}$. Following Eq. \eqref{eqq:ueven}, we have,
\begin{eqnarray}
\nonumber	u_{\text{even}} = \sum_{\substack{n_1,\ldots,n_M=0  \\ n_1+\ldots+n_M=\text{even}}}^\infty |\langle n_1 | c\rangle |^2\ldots |\langle n_M | c\rangle |^2,
\end{eqnarray} 
which can be rewritten as,
\begin{eqnarray}
\nonumber	u_{\text{even}} &=&\sum_{n_1,\ldots,n_M=0}^\infty |\langle n_1 | c\rangle |^2\ldots |\langle n_M | c\rangle |^2\\
\label{eqq:ueven2}	&&\times\frac{1+(-1)^{n_1+\ldots+n_M}}{2}.
\end{eqnarray}
By separating the fraction of Eq. \eqref{eqq:ueven2} and using the fact that state $|c\rangle^{\otimes M}$ is normalized we get,
\begin{eqnarray}
\nonumber	u_{\text{even}} &=&\frac{1}{2}+\frac{1}{2}\sum_{n_1,\ldots,n_M=0}^\infty |\langle n_1 | c\rangle |^2\ldots |\langle n_M | c\rangle |^2\\
\label{eqq:ueven3}	&&\times(-1)^{n_1+\ldots+n_M}.
\end{eqnarray}
Using Eq. \eqref{eqq:statec}, the Fock basis expansion of a coherent state $|\gamma\rangle=\exp(-|\gamma|^2/2)\sum_{n=0}^{\infty} \gamma^n/\sqrt{n!}|n\rangle$, and Eq. \eqref{eqq:normc}, we find,
\begin{eqnarray}
\nonumber u_{\text{even}} &=&\frac{1}{2}+\frac{1}{2}[e^{2|\gamma|^2}\\
\label{eqq:ueven4}&&-2 (|b_1|^2+|b_2|^2)\sinh (2|\gamma|^2)]^M,
\end{eqnarray}
and using Eq. \eqref{eqq:1minusuEven} we get,
\begin{eqnarray}
\nonumber u_{\text{odd}} &=&\frac{1}{2}-\frac{1}{2}[e^{2|\gamma|^2}\\
\label{eqq:uodd4}&&-2 (|b_1|^2+|b_2|^2)\sinh (2|\gamma|^2)]^M.
\end{eqnarray}
We observe that $u_{\text{even}}$ and $u_{\text{odd}}$ for the state of Eq. \eqref{eqq:statec}, depend only on the absolute values of the the state's coefficients when expressed as a coherent states' superposition. We note that for our $N$-mode Gaussian state, the $M$-mode produced state, and the $M$ target states, we allow $N$ and $M,\ 1\leq M<N$, to be arbitrary.  

As applications, we will consider the following target states,
\begin{eqnarray}
\label{eqq:0bar}	|\bar{0}\rangle&=&\frac{1}{N_0} (|\gamma\rangle+|-\gamma\rangle),\\
\label{eqq:1bar}	|\bar{1}\rangle&=&\frac{1}{N_1} (|\gamma\rangle-|-\gamma\rangle)\\
\label{eqq:+}	|+\rangle&=&\frac{1}{\sqrt{2}} (|\bar{0}\rangle+|\bar{1}\rangle),\\
\label{eqq:-}	|-\rangle&=&\frac{1}{\sqrt{2}} (|\bar{0}\rangle-|\bar{1}\rangle),
\end{eqnarray}
where $N_{k}=\sqrt{2[ +(-1)^k e^{-2|\gamma|^2}]},\ k=0,1$. The states of Eqs. \eqref{eqq:0bar}, \eqref{eqq:1bar}, \eqref{eqq:+}, \eqref{eqq:-}, are the computational- and rotated-basis qubit states corresponding to the coherent cat-basis qubit---one of the leading qubit candidates for all-photonic quantum computing. The states $|\bar{0}\rangle$ and $|\bar{1}\rangle$ form the so-called logical qubit basis, while the states $|+\rangle$ and $|-\rangle$ are derived by the action of a Hadamard gate (defined on the qubit basis) on the logical qubit basis' kets. For the $|\bar{0}\rangle,\ |\bar{1}\rangle$ states we find,
\begin{eqnarray}
\label{eqq:uevenstate0}	u_{\text{even}}^{|0\rangle^{\otimes M}} &=&  1,\\
\label{eqq:uoddstate0}	u_{\text{odd}}^{|0\rangle^{\otimes M}} &=&  0,\\
\label{eqq:uevenstate1}	u_{\text{even}}^{|1\rangle^{\otimes M}} &=&  \frac{1+(-1)^M}{2},\\
\label{eqq:uoddstate1}	u_{\text{odd}}^{|1\rangle^{\otimes M}} &=&  \frac{1-(-1)^M}{2}.\\
\end{eqnarray}
Consistent with the parity of the $|\bar{0}\rangle,\ |\bar{1}\rangle$ states, we see that a PNR pattern whose summation is odd cannot herald the state $|\bar{0}\rangle$, while the state $|\bar{0}\rangle$ is not impossible to be engineered if the summation of the PNR pattern is an even number. Also, if $M$ is an odd (even) number, $|\bar{1}\rangle$ cannot be heralded if the PNR pattern is summed to an even (odd) number. We note that an upper bound equal to $1$ does not mean that the state can be engineered with perfect fidelity. However, high fidelity for generating approximations of the $|\bar{0}\rangle,\ |\bar{1}\rangle$ states has been found in the literature \cite{Dakna1997,Gagatsos2019,pizzimenti2021} using the partial PNR method, even with the resource Gaussian state being zero-mean.

For the $|+\rangle,\ |-\rangle$ states we find,
\begin{eqnarray}
\label{eqq:uevenstate+}	u_{\text{even}}^{|+\rangle^{\otimes M}} &=&  \frac{1}{2},\\
\label{eqq:uoddstate+}	u_{\text{odd}}^{|+\rangle^{\otimes M}} &=&  \frac{1}{2},\\
\label{eqq:uevenstate-}	u_{\text{even}}^{|-\rangle^{\otimes M}} &=&  \frac{1}{2},\\
\label{eqq:uoddstate-}	u_{\text{odd}}^{|-\rangle^{\otimes M}} &=&  \frac{1}{2}.\\
\end{eqnarray}
Since $|+\rangle, |-\rangle$ are not zero-mean states, we expect that the upper bound should reflect that by being less than $1$. In fact, the upper bounds are low enough to conclude that the $|+\rangle, |-\rangle$ states {\em cannot} be heralded no matter what the summation of the PNR pattern is. Let us assume that we can engineer the $|\bar{0}\rangle$ or $|\bar{1}\rangle$ state with perfect fidelity from a zero-mean Gaussian state using partial PNR. Then, since it is impossible to engineer the $|+\rangle, |-\rangle$ states utilizing a zero-mean Gaussian state, we conclude that any optical implementation of a Hadamard gate (defined in the qubit space) based on Gaussian resources and partial PNR, must necessarily include displacements, in accordance with the setups presented in \cite{Ralph2003}.

\section{Fidelity upper bounds for coherent GHZ and cluster states}\label{Sec:Entangled}
Consider a non-Gaussian target state that is the multi-mode superposition,
\begin{eqnarray}
	\label{eqq:mutlimodeState}|C\rangle= \sum_{l=1}^{2^M} b_l |\bm{\gamma}^{(l)}\rangle,
\end{eqnarray}
where $b_l\in\ \mathbb{C}$ are such that state $|C\rangle$ is normalized and $|\bm{\gamma}^{(l)}\rangle$ is a product of $M$ coherent states $|\gamma\rangle$, $|-\gamma\rangle$, or any combination thereof (there exist $2^M$ such product states). We can rewrite Eq. \eqref{eqq:mutlimodeState} as,
\begin{eqnarray}
\label{eqq:mutlimodeState2}|C\rangle= \sum_{l=1}^{2^M} b_l (-1)^{\bm{\nu}(l)\cdot \hat{\bm{n}}} |\bm{\gamma}\rangle,
\end{eqnarray}
where $ |\bm{\gamma}\rangle \equiv |\bm{\gamma}^{(1)}\rangle$ is a product of $M$  coherent states $|\gamma\rangle$, $\bm{\nu}(l)$ is a vector consisting of $M$ components which are combinations of $\pm 1$ (e.g., $\bm{\nu}(1)=(1,\ldots,1)$, $\bm{\nu}(2)=(-1,1,\ldots)$, $\bm{\nu}(3)=(1,-1,1,\ldots,1)$, there exist $2^M$ such vectors), and $\hat{\bm{n}}=\left(\hat{n}_1,\ldots,\hat{n}_M\right)$.
Equation \eqref{eqq:ueven} gives,
\begin{eqnarray}
\nonumber	u_{\text{even}} &=& \sum_{\substack{n_1,\ldots,n_M=0  \\ n_1+\ldots+n_M=\text{even}}}^\infty |\langle n_1 \ldots n_M|C\rangle |^2=\\
\nonumber  &=& \sum_{n_1,\ldots,n_M=0 }^\infty |\langle n_1 \ldots n_M|C\rangle |^2 \\
\label{eqq:uevenC1} && \times\left(\frac{1+(-1)^{n_1+\ldots+n_M}}{2}\right).
\end{eqnarray}
Working out Eq. \eqref{eqq:uevenC1} and using \eqref{eqq:mutlimodeState2} we get,
\begin{eqnarray}
\nonumber	u_{\text{even}} &=& \frac{1}{2}+\frac{e^{-M |\gamma|^2}}{2} \sum_{n_1,\ldots,n_M=0 }^{\infty} \frac{|\gamma|^{2 (n_1+\ldots+n_M)}}{n_1!\ldots n_M!}\\
\label{eqq:ueven5}&& \times \Bigg|\sum_{l=1}^{2^M} b_l (-1)^{\bm{\nu}(l)\cdot \bm{n}}\Bigg|^2(-1)^{n_1+\ldots+n_M},
\end{eqnarray}
where $\bm{n}=(n_1,\ldots,n_M)$. It is hard to write Eq. \eqref{eqq:ueven5} in closed form, however if one specifies the coefficients $b_l$, the summation is rendered computable. One could write a similar to Eq. \eqref{eqq:ueven5} expression for a state like \eqref{eqq:mutlimodeState} but with different coherent amplitudes per mode, however states with equal coherent amplitudes are relevant to quantum computing. We remind the reader that the upper bound $u_{\text{odd}}$ is always given by $u_{\text{odd}}=1-u_{\text{even}}$ as per Eq. \eqref{eqq:1minusuEven}.
\subsection{GHZ states}
Let us examine the following GHZ states,
\begin{eqnarray}
	|\text{GHZ}_{\pm}\rangle =\frac{1}{N_{\pm}}\left(|\bm{\gamma}\rangle\pm|-\bm{\gamma}\rangle\right),
\end{eqnarray}
where $N_{\pm}=2\pm 2 e^{-2 M |\gamma|^2}$ are the normalization constants. Applying Eq. \eqref{eqq:ueven5} and \eqref{eqq:1minusuEven} we find,
\begin{eqnarray}
u_{\text{even}}^{|\text{GHZ}_+\rangle}&=&1,\\
u_{\text{odd}}^{|\text{GHZ}_+\rangle}&=&0,\\
u_{\text{even}}^{|\text{GHZ}_-\rangle}&=&0,\\
u_{\text{odd}}^{|\text{GHZ}_-\rangle}&=&1,
\end{eqnarray}
which is again consistent with the fact that the mean filed amplitude of GHZ states is zero and with the parity of the PNR patterrn imposed by the absence of displacement in the resource Gaussian state. We note again, that we do not prove that our upper bound is attainable. However, it has been shown that GHZ states can be produced with high fidelity \cite{Gagatsos2019} even with zero-mean resource Gaussian states.
\subsection{Coherent Cat-basis Cluster State}
Let us move to a more interesting case. Consider the $CZ_{2}$ gate whose action is defined as $CZ_{2} |\bar{0}\bar{0}\rangle= |\bar{0}\bar{0}\rangle$, $CZ_{2} |\bar{0}\bar{1}\rangle= |\bar{0}\bar{1}\rangle$, $CZ_{2} |\bar{1}\bar{0}\rangle= |\bar{1}\bar{0}\rangle$, and $CZ_{2} |\bar{1}\bar{1}\rangle= -|\bar{1}\bar{1}\rangle$, and therefore is an entangling operation when it acts on $|++\rangle$. In this work, we denote as $CZ$ any product consisting of multiple two-mode $CZ_{2}$ gates, acting on any two qubits of a multi-qubit product state. In fact, we consider that $CZ$  acts on the state $|+\rangle^{\otimes M}$, i.e., $CZ |+\rangle^{\otimes M}$, to create entanglement between any possible couple of $|+\rangle$ states at the same time, therefore creating a cluster state on the coherent-cat basis, i.e,
\begin{eqnarray}
\label{eqq:stateCCCS}	|\text{CCCS}\rangle = CZ |+\rangle^{\otimes M},
\end{eqnarray} 
is any coherent cat-basis cluster state. Using Eq. \eqref{eqq:+}, said state can be written as,
\begin{align}
\nonumber &	|\text{CCCS}\rangle = \frac{1}{2^{M/2}} \Bigg[s_1 |\bar{0}\ldots \bar{0}\rangle\\
\nonumber &+ s_2 |\bar{1} \bar{0} \ldots \bar{0}\rangle+\ldots + s_{M+2} |\bar{0}  \ldots \bar{0}\bar{1}\rangle+\\
& \nonumber +s_{M+3} |\bar{1} \bar{1} \bar{0} \rangle +\ldots+ s_{\binom{M}{2}-(M+3)} |\bar{0}\ldots \bar{1}\bar{1}\rangle\\
& \nonumber+\ldots\\
&\label{eqq:stateCCCS2} +s_{2^M} |\bar{1}\ldots\bar{1}\rangle\Bigg].
\end{align}
Let us explain the terms of Eq. \eqref{eqq:stateCCCS2}: The fist line is the product state $|\bar{0}\rangle^{\otimes M}$ and there is $\binom{M}{0}=1$ such state. The second line is a product of $M-1$ $|\bar{0}\rangle$ states and $1$ $|\bar{1}\rangle$ state which can take any position and therefore there are $\binom{M}{1}=M$ such states present in said line. Similarly, in the third line the $2$ $|\bar{1}\rangle$ states can take any position and there are $\binom{M}{2}$ such states, and so on. Finally, in the last line all states are $|\bar{1}\rangle$ and there is only one such state as $\binom{M}{M}=1$. The prefactors $s_i,\ i=1,\ldots,2^M$ can only be $\pm 1$, according to the prescription of any given $CZ$ gate creating any desired cluster configuration. In fact, unless there are at least two $|\bar{1}\rangle$ vectors present, one can set $s_i=1,\ \forall i$.

From Eqs. \eqref{eqq:ueven} and \eqref{eqq:stateCCCS2} we have,
\begin{align}
\nonumber &	u_{\text{even}}^{|\text{CCCS}\rangle} = \frac{1}{2^{M}} \sum_{\substack{n_1,\ldots,n_M=0  \\ n_1+\ldots+n_M=\text{even}}}^\infty \Big|s_1 \langle n_1 \ldots n_M |\bar{0}\ldots \bar{0}\rangle+\ldots\\
&\label{eqq:uevenCCCS1} +s_{2^M} \langle n_1 \ldots n_M |\bar{1}\ldots\bar{1}\rangle\Big|^2.
\end{align}
Let us examine the cross-terms of the expansion of $|\ldots|^2$ in Eq. \eqref{eqq:uevenCCCS1}, i.e., terms with different prefactors $s_i$. Any such term is the product $\langle n_1\ldots n_M| \bar{q}_1\ldots \bar{q}_M\rangle \langle \bar{q}'_1\ldots \bar{q}'_M|n_1\ldots n_M\rangle$, with $\bar{q}_1,\ldots,\bar{q}_M =0,1$ and $\bar{q}'_1,\ldots,\bar{q}'_M =0,1$. Since they are cross-terms, there is at least one $i=1,\ldots,M$ such that $\bar{q}_i\neq \bar{q}'_i$. Given the Fock expansion coefficients $\langle n|\bar{0}\rangle = N_0^{-1}e^{-|\gamma|^2/2}\gamma^n (1+(-1)^n)/\sqrt{n!}$ and $\langle n|\bar{1}\rangle = N_1^{-1}e^{-|\gamma|^2/2}\gamma^n (1-(-1)^n)/\sqrt{n!}$, we see that any cross-term will be proportional to $(1+(-1)^{n_i})(1-(-1)^{n_i})=1-(-1)^{2n_i}=0$, for at least one $i$. 

Therefore, the only non-zero terms in Eq. \eqref{eqq:uevenCCCS1} are of the form $|s_i \langle n_1\ldots n_M| \bar{q}_1\ldots \bar{q}_M\rangle|^2=|\langle n_1\ldots n_M| \bar{q}_1\ldots \bar{q}_M\rangle|^2$, since $|s_i|^2=1$. Therefore, we have,
\begin{eqnarray}
\label{eqq:uevenCCCS2}	u_{\text{even}}^{|\text{CCCS}\rangle} = \frac{1}{2^M}\sum_{k=0}^M \binom{M}{k} u_{\text{even}}^{|\bar{0}^{\otimes (M-k)} \bar{1}^{\otimes k}\rangle}.
\end{eqnarray}
 From Eq. \eqref{eqq:ueven} and following the methods of Section \ref{Sec:Product}, we find that,
 \begin{eqnarray}
 \label{eqq:ueven0Mk1k}	u_{\text{even}}^{|\bar{0}^{\otimes (M-k)} \bar{1}^{\otimes k}\rangle}=\frac{1+(-1)^k}{2}.
 \end{eqnarray}
 Finally, from Eqs. \eqref{eqq:uevenCCCS2}, \eqref{eqq:ueven0Mk1k}, and \eqref{eqq:1minusuEven} we find,
 \begin{eqnarray}
 	\label{eqq:uevenCCCS3}	u_{\text{even}}^{|\text{CCCS}\rangle} &=& \frac{1}{2},\\
 	 	\label{eqq:uoddCCCS3}	u_{\text{odd}}^{|\text{CCCS}\rangle} &=& \frac{1}{2}.
 \end{eqnarray}
Any $|\text{CCCS}\rangle$ has inherently non-zero mean-field amplitude because the $|+\rangle$ has non-zero displacement. The upper bound of Eqs. \eqref{eqq:uevenCCCS3} and \eqref{eqq:uoddCCCS3} quantifies the damage of not considering displacement as a resource.  An upper bound equal to $1/2$ on the fidelity with any produced state shows that any $|\text{CCCS}\rangle$ state is well beyond reach with a zero-mean Gaussian resource state. 

\section{Conclusions and discussion}\label{Sec:Conclusions}
Partial PNR is the new trend for non-Gaussian bosonic state engineering because essentially it circumvents the technical difficulties of constructing non-Gaussian optical unitary operations. However, the are two main drawbacks in said approach: (i) optimization methods are needed to reveal an optimal resource Gaussian state that maximizes the fidelity and probability of occurrence of an acceptable produced state (ii) fidelity is merely a necessary criterion. Any numerical optimization typically does not give intuition on the underlying physics of state engineering. In this paper we asked what would happen if we forbid our resource state to posses any displacements and therefore reveal the implications on coherent-cat basis clusters under any optimization of such resource. We recognized that zero displacement restricts the parity of the observed PNR pattern and therefore it restricts the Fock expansion coefficients (modulo squared) one should sum up to derive a fidelity upper bound, yielding a hard $1/2$ upper bound for target states with non-zero mean field amplitude such as the $|+\rangle$, $|-\rangle$, and $|\text{CCCS}\rangle$ states.

As a byproduct, we argued that any optical implementation, i.e., based on Gaussian resources and partial PNR, of a qubit Hadamard gate (an operation transforming $|\bar{0}\rangle \rightarrow |+\rangle$) must necessarily include displacements. This Hadamard gate could be a separate primitive consisting $m$ displaced squeezed states as inputs to a passive $n$-mode interferometer ($m<n$). The rest of the $n-m$ input modes could be the output state of another partial PNR based scheme which produces the $|\bar{0}\rangle$ or $|\bar{1}\rangle$ states. The Hadamard optical primitive and the $|\bar{0}\rangle$ or $|\bar{1}\rangle$ state generator could be combined into a single interferometer, with single mode dispalced squeezed inputs, and an array of PNR detectors at the output, some of which control the production of $|\bar{0}\rangle$ or $|\bar{1}\rangle$ and another PNR subset the realization of the Hadamard gate.

It is known that PNR detectors and Gaussian states comprise a universal resource set \cite{Braun1999}. Therefore, by working with a general pure Gaussian state, i.e., including displacements, universality must be restored. Apparently, a displacement $D(\alpha)$ on the undetected output would not suffice as it can be easily seen that for example $D(\alpha)|\bar{0}\rangle \neq |+\rangle$. All displacements must be applied on the squeezed single mode states going into the passive interferometer, or just before partial PNR (i.e. equivalently partially projecting a zero-mean Gaussian state onto displaced Fock states). However, a constructive way of designing partial PNR based schemes which would attain universality is still elusive.

The holy grail of this line of research would be a systematic theory for non-Gaussian state engineering for specific classes of states that are useful in various quantum information processing tasks such as cluster states for quantum computing, all-optical quantum repeaters, metrologically-optimal states in distributed quantum sensing, etc. One specific interesting question that relates to the states considered in this paper is: Whether the GHZ states considered in this work can be transformed into the CCCS by using local unitaries (e.g., it is known that a star-topology cluster state and a GHZ state are local-Hadamard equivalent), where the local unitaries are themselves realized by post-selected non-Gaussian ancilla states which in turn were heralded using Gaussian states and PNR detectors \cite{Ralph2003}. 

Such questions could be answered by expanding the mathematical formalism developed in \cite{Gagatsos2019,pizzimenti2021} (also given as App. \ref{App} of this work) to include displacements. This could catalyze further progress toward the non-Gaussian state engineering, if not in providing constructive ways for attaining universality, but at least for constructing optical implementations for specific useful to quantum computation primitives. 

\begin{acknowledgments}
CNG was supported by Xanadu Quantum Technologies. CNG and SG acknowledge Xanadu Quantum Technologies for supporting multiple useful discussions on this topic. CNG also acknowledges partial funding support from the National Science Foundation, award number 2122337.
\end{acknowledgments}

\bibliography{FidPhotonSubBIB.bib}

\newpage
\appendix
\section{Derivations of probabilities and Fock coefficients of engineered non-Gaussian states}\label{App}
\renewcommand{\thesubsection}{\arabic{subsection}}
\def\theequation{A\arabic{equation}}
\setcounter{equation}{0}
\renewcommand{\thefigure}{A\arabic{figure}}    
\setcounter{figure}{0}
The following can be found as Section II of \cite{pizzimenti2021} co-authored by CNG, SK, and collaborators. Said subsection was authored by CNG. 

Here we briefly review the results in Ref.~\cite{Gagatsos2019} and then evolve those to further worked-out formulas. Among other things, in Ref.~\cite{Gagatsos2019} it was proven that any $N$-mode pure Gaussian state $|\Psi\rangle$ with covariance matrix (CM) $V$ and displacement vector $\vec{x}_\beta$ can be written in the coherent basis $|\vec{\alpha}\rangle$ as
\begin{eqnarray}
\label{eq:PsiWithDisplacements}|\Psi\rangle = \int d^{2N} \vec{x}_\alpha\ K(\vec{x}_\alpha) |\vec{\alpha} \rangle,
\end{eqnarray}
where
\begin{eqnarray}
\label{eq:Kfunction} K(\vec{x}_{\alpha})&=&\frac{e^{-\frac{1}{2} (\vec{x}_{\alpha}-\vec{x}_{\beta})^T \mathcal{B} (\vec{x}_{\alpha}-\vec{x}_{\beta})+\frac{1}{2} \vec{x}_\alpha^T \mathcal{Y} \vec{x}_\beta }}{(2\pi)^N(\det \Gamma)^{1/4}},
\end{eqnarray}
with $\Gamma=V+I/2$,
\begin{flalign}
\label{eq:Bmatrix}	\mathcal{B}=\frac{1}{2}
\begin{pmatrix}
A + \frac{i}{2}\left(C+C^T\right) & C - \frac{i}{2}\left(A-B\right) \\
C^T - \frac{i}{2}\left(A-B\right) & B - \frac{i}{2}\left(C+C^T\right)
\end{pmatrix},&&\\
\mathcal{Y}=
\begin{pmatrix}
0 & iI \\
-iI& 0
\end{pmatrix},&&
\end{flalign}
where $A=A^T$, $B=B^T$, and $C$ are defined as the blocks of $\Gamma^{-1}$ as follows:
\begin{eqnarray}
\label{eq:GammaInv}	\Gamma^{-1}=
\begin{pmatrix}
A & C\\
C^T & B
\end{pmatrix}.
\end{eqnarray}
Note that we have simplified the expressions compared to Ref.~\cite{Gagatsos2019}.
We note that since the CM $V$ is symmetric, $\Gamma$ and $\Gamma^{-1}$ are also symmetric.
We work with the convention $\hbar=1$ (therefore the CM of vacuum is $I/2$) and consider the $qqpp$ representation where vectors are defined as  ${\vec{x}_{\alpha}^T=(\vec{q}_{\alpha}^T,\vec{p}_{\alpha}^T)}$ 
with $\vec{q}^T_{\alpha}=(q_{\alpha_1},\ldots,q_{\alpha_N})$ and $\vec{p}_{\alpha}^T=(p_{\alpha_1},\ldots,p_{\alpha_N})$ the canonical position and momentum vectors.
The volume element for integration is then defined as $d^{2N} \vec{x}_{\alpha}=dq_{\alpha_1}\ldots dq_{\alpha_N} dp_{\alpha_1}\ldots dp_{\alpha_N}$, and $\alpha_i=(q_{\alpha_i}+i p_{\alpha_i})/\sqrt{2}$.

The coherent basis representation is a valuable tool for working on photon-subtraction-based or, more generally, partial PNR detection schemes aimed at engineering Gaussian states into desired non-Gaussian states. Photon subtraction can be modeled either (i) as a beam-splitter whose two input ports are fed with the $i$th mode of $|\Psi\rangle$ and vacuum $|0\rangle$, respectively, followed by PNR detection on the lower output port; or (ii) simply by acting the annihilation operator $\hat{a}_i$, where the index $i$ refers to the mode, on $|\Psi\rangle$. Therefore, the photon subtraction operator will act only on the basis vectors of the state, i.e., coherent states in this instance. The action of beam-splitters or annihilation operators on coherent states is straightforward, making this basis particularly efficient for analytical or numerical evaluation. 
The situation is similar for partial PNR detection on a Gaussian state written as a coherent state expansion; 
the projection of a coherent state on a Fock state is the well known expression $\langle n|\alpha \rangle = \exp(-|\alpha|^2/2)\alpha^n/\sqrt{n!}$.

In Ref.~\cite{Gagatsos2019} it was shown that the probability of a length-$N$ PNR pattern
for an $N$-mode Gaussian state $|\Psi\rangle$ with zero displacements, i.e., $\vec{x}_\beta=0$ in Eq.~\eqref{eq:PsiWithDisplacements}, is given by
\begin{flalign}
\nonumber   P_{n_1 \ldots n_N}=|\langle n_1 \ldots n_N| \Psi\rangle|^2 &&\\
\label{eq:GBSprob} =  \frac{1}{\det \mathcal{H} \sqrt{\det\Gamma} \prod\limits_{i=1}^{N} n_i!2^{n_i}}\big|\mathcal{I}_{n_1 \ldots n_N}\big|^2,&&
\end{flalign}
where
\begin{flalign}
\label{eq:IN}	\mathcal{I}_{n_1 \ldots n_N} =\int d^{2N}\vec{x}_\alpha R(\vec{x}_{\alpha})  \prod\limits_{i=1}^{N}(q_{\alpha_i}+i p_{\alpha_i})^{n_i},&&\\
\label{eq:Rdistr}	R(\vec{x}_{\alpha}) =  \frac{\sqrt{\det \mathcal{H}}}{(2\pi)^{N}}  e^{-\frac{1}{2}\vec{x}_\alpha^T \mathcal{H} \vec{x}_\alpha},&&
\end{flalign}
and $\mathcal{H}=\mathcal{B}+I/2$. Equation \eqref{eq:IN} can be rewritten as
\begin{eqnarray}
\label{eq:Hafnian}	\mathcal{I}_{n_1 \ldots n_N} = \left\{
\begin{array}{ll}
0&\Sigma=\textrm{odd},\\
\textrm{Hf}\left(\sigma\right)&\Sigma= \textrm{even},
\end{array}
\right.
\end{eqnarray}
where $\Sigma=\sum_{i=1}^{N}n_i$, $\textrm{Hf}\left(\sigma\right)$ is the \emph{loop hafnian} (to be briefly explained in Sec.~\ref{sec:FandHinv}) of the matrix $\sigma$ with elements $\sigma_{ij}=\langle s_i s_j \rangle$, where $1\leq i,j\leq \Sigma$ and  $s_i=q_{\alpha_i}+ip_{\alpha_i}$. The hafnian in Eq. \eqref{eq:Hafnian} represents the mean value $\langle s_1^{n_1}\ldots s_N^{n_N} \rangle$ under the Gaussian distribution of Eq.~\eqref{eq:Rdistr}.

In this work, we will derive the explicit relation of the matrix $\sigma$ to the matrix $\mathcal{H}^{-1}$ and consequently to matrices $\Gamma$ and $V$. We also give the expressions for the Fock expansion coefficients of the produced non-Gaussian states and simplify further the expressions. The following subsections summarize new simplifications, observations, and new results which improve on Eqs. (\ref{eq:GammaInv}--\ref{eq:Rdistr}).

\subsection{The determinant and inverse of $\Gamma$}\label{sec:Gamma}
The matrix $\Gamma$ is defined as $\Gamma=V+I/2$, where $V$ is the CM and $I$ the identity matrix. Since $V$ corresponds to a pure Gaussian state, it can be written as $V=S_p V_0 S_p^T$, where $S_p$ is an orthogonal symplectic matrix for a general passive transformation (beam-splitters and phase rotations, but not squeezers) and $V_0$ is the CM for a product of $N$ single mode squeezed vacuum states, i.e., the diagonal matrix
\begin{flalign}
V_0=\frac{1}{2}\textrm{diag} \left( e^{2 r_1},\ldots,e^{2 r_N}, e^{-2 r_1},\ldots, e^{-2 r_N}\right),
\end{flalign}
where $r_1,\ldots,r_N$ are the real and positive squeezing parameters for each of the $N$ single-mode squeezed vacuum states (note that the phase of the squeezing has been absorbed into the orthogonal symplectic transformation $S_p$).

We have the following relation,
\begin{eqnarray}
\det\Gamma&=&\det\left[S_p\left(V_0+\frac{I}{2}\right)S_p^T\right]\\
&=&\det S_p \det\left(V_0+\frac{I}{2}\right) \det S_p^T,
\end{eqnarray}
from which we write
\begin{eqnarray}
\label{eq:detGamma1}\det\Gamma= \det\left(V_0+\frac{I}{2}\right)
\end{eqnarray}
since $\det S_p=\det S_p^T=1$ as both $S_p$ and $S_p^T$ are symplectic matrices.
The right hand side of Eq. \eqref{eq:detGamma1} is the determinant of a diagonal matrix from which we find
\begin{eqnarray}
\label{eq:detGamma2}\det\Gamma = \prod_{i=1}^N \cosh^2 r_i.
\end{eqnarray}
Therefore, Eq. \eqref{eq:GBSprob} is rewritten as
\begin{eqnarray}
\label{eq:GBSprob2} P_{n_1 \ldots n_N}=\frac{\big|\mathcal{I}_{n_1\ldots n_N}\big|^2}{\det \mathcal{H} \prod\limits_{i=1}^{N} n_i!2^{n_i} \cosh r_i}.
\end{eqnarray}
In the case where the input squeezing is the same among all single mode squeezed vacuum states, i.e. $r_1=\ldots=r_N=r$, Eq. \eqref{eq:detGamma2} reduces to $\det\Gamma=\cosh^{2N}r$.

Now let us simplify Eq. \eqref{eq:GammaInv}. We can write $\Gamma=S_p (V_0+I/2)S_p^T$, and since $S_p^{T^{-1}}=S_p$ is a symplectic orthogonal matrix we have
\begin{eqnarray}
\label{eq:GammaInv2} \Gamma^{-1} = S_p \left(V_0+\frac{1}{2}\right)^{-1} S_p^T.
\end{eqnarray}
The symplectic orthogonal matrix $S_p$ has the following block matrix structure and properties:
\begin{eqnarray}
\label{eq:Mp}S_p &=& \begin{pmatrix}
S_A & S_B\\
-S_B & S_A
\end{pmatrix}\\
\label{eq:MpConstr1}&& S_A^T S_B = S_B^T S_A,\\
\label{eq:MpConstr2}&& S_A S_B^T = S_B S_A^T,\\
\label{eq:MpConstr3}&& S_A^T S_A+S_B^T S_B = I,\\
\label{eq:MpConstr4}&& S_A S_A^T+S_B S_B^T = I.
\end{eqnarray}
Moreover, since $V_0$ is diagonal we can write
\begin{eqnarray}
\label{eq:invGamma_0}
\left(V_0+\frac{1}{2}\right)^{-1} = I+\begin{pmatrix}
-T & 0\\
0 & T
\end{pmatrix},
\end{eqnarray}
where $T=\textrm{diag}\left(\tanh r_1,\ldots,\tanh r_N\right)$. In virtue of Eqs. \eqref{eq:GammaInv2}, \eqref{eq:Mp}, and \eqref{eq:MpConstr2}, we find that in Eq. \eqref{eq:GammaInv} 
\begin{eqnarray}
\label{eq:blockA} A &=&-S_A T S_A^T + S_B T S_B^T,\\
\label{eq:blockC} C&=&C^T= S_A T S_B^T+S_B T S_A^T,\\
&& A+B= 2I.
\end{eqnarray}
Therefore, in the most general case possible, Eq. \eqref{eq:GammaInv} is simplified to
\begin{eqnarray}
\label{eq:GammaInv3} \Gamma^{-1} = \begin{pmatrix}
A & C \\
C & 2 I -A,
\end{pmatrix}
\end{eqnarray}
where $A$ and $C$ are given in Eqs. \eqref{eq:blockA} and \eqref{eq:blockC}, respectively, as functions of the passive symplectic transformation $S_p$ and the input squeezing parameters.

Consequently, matrix $\mathcal{B}$ of Eq. \eqref{eq:Bmatrix} simplifies to
\begin{eqnarray}
\label{eq:Bmatrix2}\mathcal{B}=\frac{1}{2}\begin{pmatrix}
A+ i C & C-i(A-I) \\
C-i(A-I) & 2I-A-iC
\end{pmatrix}.
\end{eqnarray}
\subsection{The determinant and inverse of $\mathcal{H}$}\label{sec:H}
The matrix $\mathcal{H}$ appearing in Eq.~\eqref{eq:Rdistr} is defined as 
\begin{eqnarray}
\label{eq:Hmatrix}\mathcal{H}=\mathcal{B}+I/2. 
\end{eqnarray}
We find it easier if we transform as $\tilde{\mathcal{H}}= W^\dagger\mathcal{H}W$ using the unitary matrix $W$ 
defined as
\begin{eqnarray}
\label{eq:Wmatrix}W=\frac{1}{\sqrt{2}}\begin{pmatrix}
I & I \\
-i I & i I
\end{pmatrix}.
\end{eqnarray}
Utilizing Eqs. \eqref{eq:Bmatrix2}, \eqref{eq:Hmatrix}, and \eqref{eq:Wmatrix} we find
\begin{eqnarray}
\label{eq:Hmatrix2}\tilde{\mathcal{H}}= \begin{pmatrix}
I & A-I +i C\\
0 & I
\end{pmatrix},
\end{eqnarray}
from which we see that $\det \tilde{\mathcal{H}}=\det I=1$. Since $|\det W|^2=1$, we have $\det \tilde{\mathcal{H}}=\det\mathcal{H}$ and conclude that
\begin{eqnarray}
\label{eq:detH}\det\mathcal{H}=1.
\end{eqnarray}
Therefore, Eqs. \eqref{eq:Rdistr} and  \eqref{eq:GBSprob2} are further simplified to
\begin{eqnarray}
\label{eq:GBSprob3} P_{n_1 \ldots n_N}&=&\frac{\big|\mathcal{I}_{n_1\ldots n_N}\big|^2}{ \prod\limits_{i=1}^{N} n_i!2^{n_i} \cosh r_i},\\
\label{eq:Rdistr2}	R(\vec{x}_{\alpha}) &=&  \frac{1}{(2\pi)^{N}}  e^{-\frac{1}{2}\vec{x}_\alpha^T \mathcal{H} \vec{x}_\alpha}.
\end{eqnarray}
Let us derive a convenient expression for $\mathcal{H}^{-1}$. Again, we work with $\tilde{\mathcal{H}}$ and observe that
\begin{eqnarray}
\label{eq:HmatrixTildeInv}\tilde{\mathcal{H}}^{-1}= \begin{pmatrix}
I & -(A-I +i C)\\
0 & I
\end{pmatrix}
\end{eqnarray}
is indeed the inverse of $\tilde{\mathcal{H}}$, i.e., it satisfies $\tilde{\mathcal{H}} \tilde{\mathcal{H}}^{-1}=I$. Since $\tilde{\mathcal{H}}= W^\dagger\mathcal{H}W$ we find that $\mathcal{H}^{-1} = W \tilde{\mathcal{H}}^{-1} W^\dagger$ and finally
\begin{flalign}
\label{eq:Hinv}\mathcal{H}^{-1}=\frac{1}{2}
\begin{pmatrix}
3 I -A-i C & i (A-I+i C)\\
i (A-I+i C) & I+A+ i C
\end{pmatrix}.
\end{flalign}
Therefore, using Eqs. \eqref{eq:blockA}, \eqref{eq:blockC}, and \eqref{eq:Hinv}, any given passive symplectic transformation $S_p$, and input squeezing parameters, one can readily write $\mathcal{H}^{-1}$---the importance of which will become apparent in the next subsections.
\subsection{The relation of matrix $\sigma$ to matrix $\mathcal{H}^{-1}$}\label{sec:FandHinv}

Making use of Eq.~\eqref{eq:Rdistr2}, we can express the matrix elements of $\sigma$ as
\begin{eqnarray}
\nonumber \sigma_{ij}&=&\langle \left(q_{\alpha_i}+i p_{\alpha_i}\right)\left(q_{\alpha_j}+i p_{\alpha_j}\right)\rangle=\\
\nonumber && \frac{1}{(2\pi)^N}\int d^{2 N} \vec{x}_{\alpha} \exp \left(-\frac{1}{2} \vec{x}_{\alpha}^{T} \mathcal{H} \vec{x}_{\alpha}\right)\\
\nonumber &&\times\left(q_{\alpha_{i}}+i p_{\alpha_{i}}\right)\left(q_{\alpha_{j}}+i p_{\alpha_{j}}\right)=\\
\label{eq:FandHinv}&& \left.\frac{d}{d \lambda_{i}} \frac{d}{d \lambda_{j}} \exp \left(\frac{1}{2} \vec{\Lambda}^{T} \mathcal{H}^{-1} \vec{\Lambda}\right)\right|_{\vec{\Lambda}=\overrightarrow{0}},
\end{eqnarray}
where $\vec{\Lambda}^T=(\vec{\lambda}^T, i \vec{\lambda}^T)$ is a $2N$-dimensional vector with $\vec{\lambda}^T=\left(\lambda_1,\ldots,\lambda_N\right)$ a real $N$-dimensional vector. Viewing $\frac{1}{2} \vec{\Lambda}^{T} \mathcal{H}^{-1} \vec{\Lambda}$ in the exponential of the right hand side of Eq. \eqref{eq:FandHinv} as a polynomial in $\lambda_i$,  Eq. \eqref{eq:FandHinv} is equal to the coefficient of $\lambda_i\lambda_j$. This way, we can write
\begin{eqnarray}
\label{eq:FandHinv2}\sigma_{ij} = 2 (\mathcal{H}^{-1}_{ij}-\mathcal{H}^{-1}_{i+N\ j+N}).
\end{eqnarray}
From the covariance matrix $V$, one can find matrix $\Gamma^{-1}$ and therefore matrix $\sigma$ using Eqs. \eqref{eq:Hinv} and \eqref{eq:FandHinv2}, which is required in the calculation in Eq. \eqref{eq:Hafnian}.

The Gaussian moment problem of Eq. \eqref{eq:IN} represents a hafnian calculation and is related to the Gaussian boson sampling paradigm \cite{Hamilton2017}. When the indices $i,j$ are equal this corresponds to a loop, i.e., matching an object with itself. Therefore, it is typically referred to as a loop hafnian. 
\subsection{Occurrence probability of any produced state}\label{sec:Probability}
\begin{figure}
	\includegraphics[width=\columnwidth]{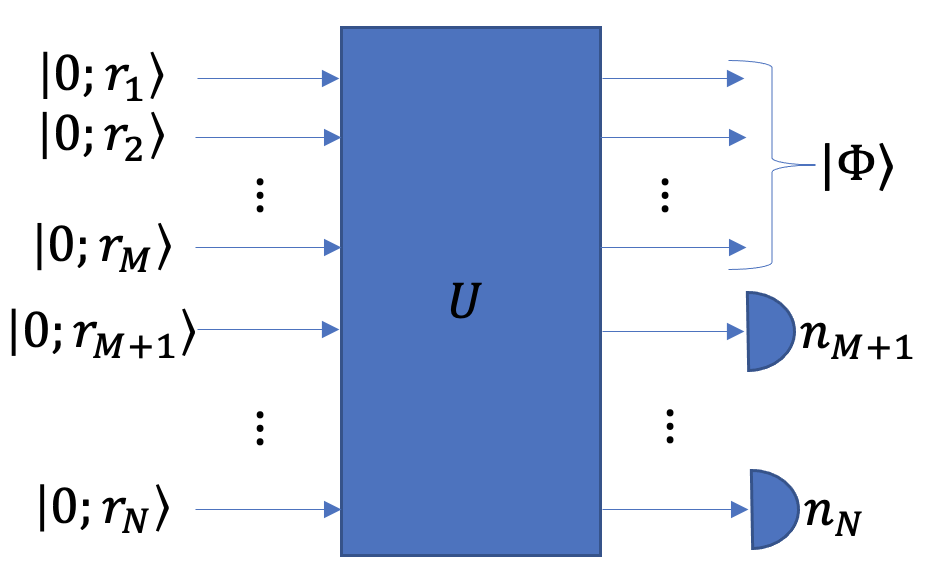}
	\caption{Concept of heralding an $M$-mode state $\ket{\Phi}$ from $N$ single-mode, zero-displacement squeezed resource states and $N\times N$ unitary operation $U$. 
		Partial PNR detection on the $N-M$ lower modes produces a non-Gaussian state on the undetected $M$ modes. 
	}
	\label{fig:GenericScheme}
\end{figure}
Equation \eqref{eq:GBSprob3} is the probability of finding $n_i$ photons in each one of the $i=1,\ldots,N$ modes. If we wish to engineer the $N$-mode Gaussian state into an $M$-mode ($M<N$) non-Gaussian one as in Fig. \ref{fig:GenericScheme}, we leave $M$ modes undetected; without loss of generality we assume the undetected modes are the $M$ upper modes. The probability of the PNR pattern $(n_{M+1}, \ldots , n_N)$ on the lower detected modes is precisely the probability $P_{n_{M+1}, \ldots ,n_N}$ of producing the corresponding non-Gaussian state. This probability is
\begin{eqnarray}
\label{eq:P}P\equiv P_{n_{M+1},\ldots,n_N} = \sum_{n_1,\ldots,n_M=0}^{\infty} P_{n_1,\ldots,n_N}.
\end{eqnarray}
For numerical simulations, the above sum must be truncated to a finite upper limit, which should be chosen with care to ensure that it encompasses all Fock coefficients of nonnegligible probability. This condition can be verified in practice by successively increasing the limits and observing no change to $P$.
\subsection{Fock expansion coefficients of the produced state}
The non-Gaussian state $|\Phi\rangle$ on the $M$ undetected modes (see Fig.~\ref{fig:GenericScheme}), can be written as a partial projection on Fock states of the detected modes:
\begin{eqnarray}
\label{eq:phi} |\Phi\rangle =\frac{1}{\sqrt{P}} \langle n_{M+1}\ldots n_N | \Psi \rangle,
\end{eqnarray}
where $P$ is given in Eq. \eqref{eq:P} and $|\Psi\rangle$ is the input $N$-mode Gaussian state. 

The Fock expansion coefficients of heralded state $|\Phi\rangle$ are $c_{n_1 \ldots n_M}=\langle n_1\ldots n_M|\Phi\rangle$. Using Eqs. \eqref{eq:IN} and \eqref{eq:phi} we find
\begin{eqnarray}
\label{eq:FockCoef}c_{n_1 \ldots n_M} = \frac{\mathcal{I}_{n_1 \ldots n_M n_{M+1}\ldots n_N}}{\sqrt{P} \prod\limits_{i=1}^{N} \sqrt{n_i!2^{n_i} \cosh r_i}},
\end{eqnarray}
where the numerator is given by Eq. \eqref{eq:IN}. Therefore, for any given partial PNR pattern $(n_{M+1},\ldots,n_N)$ one can compute the Fock expansion coefficients of the produced state $|\Phi\rangle$, which can be benchmarked against a target non-Gaussian state $\ket{\Phi_t}$ through direct comparison of Fock coefficients or collectively through fidelity $\mathcal{F}=|\langle\Phi_t|\Phi\rangle|^2$.

\end{document}